\pdfoutput=1

\documentclass[prl,showpacs,twocolumn,superscriptaddress,nofootinbib]{revtex4}

\usepackage{amsmath,amsfonts,amssymb,amstext}
\usepackage[dvipdfm]{graphicx}
\usepackage[colorlinks,linkcolor=blue]{hyperref}

\begin{document}
\title{Strongly Enhanced Spin Squeezing via Quantum Control}
\date{\today}

\author{Collin M. Trail}
\affiliation{Center for Quantum Information and Control (CQuIC)}
\affiliation{Department of Physics and Astronomy, University of New Mexico}

\author{Poul Jessen}
\affiliation{Center for Quantum Information and Control (CQuIC)}
\affiliation{College of Optical Sciences and Department of Physics, University of Arizona}

\author{Ivan Deutsch}
\affiliation{Center for Quantum Information and Control (CQuIC)}
\affiliation{Department of Physics and Astronomy, University of New Mexico}

\begin{abstract}
We describe a new approach to spin squeezing based on a double-pass Faraday interaction between an optical probe and an optically dense atomic sample. A quantum eraser is used to remove residual spin-probe entanglement, thereby realizing a single-axis twisting unitary map on the collective spin. This interaction can be phase-matched, resulting in {\em exponential} enhancement of squeezing.  In practice the scaling and peak squeezing depends on decoherence, technical loss, and noise. A simplified model indicates $\sim\!\!10$\;dB of squeezing should be achievable with current laboratory parameters.
\end{abstract}

\maketitle

The ability to control a complex quantum system is increasingly important to studies of quantum many-body physics, precision measurement, and quantum information processing.  One platform with great potential for implementing such control is the collective spin of an atomic ensemble coupled to the Stokes vector of a quantized light field \cite{hammer2008,deutsch2010}.   Given a sufficiently strong interaction, it is possible to create entanglement between the atoms and the field, and also between atoms within the ensemble through their coupling to the field, which acts as a shared Òquantum data busÓ.  The quantum correlations created between individual atomic spins can lead to squeezing of the quantum fluctuations in a quadrature of the collective spin below that of a coherent state \cite{kuzmich2000, appel2009, takahashi2009, vuletic2010, mitchell2010}.  Such squeezed states have direct applications in quantum sensing and metrology and are the foundation for quantum information processing with Òcontinuous variableÓ encoding \cite{braunstein2005}.  

In this paper we show how one can use a series of optical probe pulses in a double-pass geometry \cite{takeuchi2005} to create spin squeezing that, for a time short compared to the time scale for decoherence, improves {\em exponentially} with coupling strength.  This represents a significant improvement over existing schemes for which spin squeezing scales roughly linearly with coupling strength \cite{hammer2008}.  The key is to achieve quantum coherent control of the collective spin.  One can accomplish this starting from a proposal by Takeuchi {\em et al} \cite{takeuchi2005}.   In this protocol the polarization of a probe pulse is correlated with the spin through Faraday rotation during a first pass, and then reflected back through the ensemble for a second pass, where it acts as a fictitious magnetic field that produces a spin-dependent (and thus nonlinear) rotation of the spin.  Residual entanglement between polarization and spin after the second pass leads to decoherence and excess noise on the spin when the light pulse is discarded, but even so it is still possible to achieve a limited degree of squeezing in one spin quadrature.  We propose an improved protocol wherein the quantum information carried by the probe is removed by a quantum eraser \cite{scully1981}, resulting in a purely unitary evolution of the collective spin.  In this situation, appropriate control with an applied magnetic field allows the squeezing to be phase matched in a manner analogous to squeezing of optical fields.  The result is a reduction in quantum projection noise that scales exponentially with the coupling constant, and a commensurate increase in quantum correlations.  Related multi-pass scenarios have been considered previously that in principle create an exponential amount of two-mode squeezing in the entanglement between atoms and light \cite{hammer2004}, but none of these lead to exponential growth of the spin squeezing of atoms alone.  The maximum degree of squeezing and its scaling with coupling strength will ultimately depend on decoherence and noise.  Our preliminary model shows that $\sim\!\!10$\;dB of squeezing should be possible in the presence of realistic levels of photon scattering, optical pumping, optical losses and detector noise,  for optimistic but not unreasonable coupling strengths such as might be achieved with atomic samples in optical dipole traps \cite{mitchell2009}. 

As in previous works, collective spin control can be achieved based on the Faraday interaction between the collective atomic spin vector $\mathbf{J}$ and the photonic Stokes vector $\mathbf{S}$, described by a unitary entangling operator $U_F = e^{-i \chi J_z S_3}$ \cite{hammer2008, deutsch2010}. The characteristic Faraday rotation angle per unit spin angular momentum is $\chi = (\sigma_0 / A) (\Gamma / 3 \Delta) $, where $\sigma_0 = 3 \lambda^2 / 2 \pi$  is the resonant scattering cross-section for unit oscillator strength, $\lambda$ is the transition wavelength, $\Gamma$ is the atomic linewidth, $\Delta$ is the detuning from resonance, and $A$ is the cross sectional area of the light spatial mode.  Under the usual Holstein-Primakov approximation (HPA) \cite{madsen2004}, for a large number atoms, $N_A$, and photons in the probe light pulse, $N_L$, we define canonical variables for Gaussian fluctuations about the mean field, $X_A \equiv J_y/\sqrt{N_A/2}$, $P_A \equiv J_z/\sqrt{N_A/2}$, $X_L \equiv S_2/\sqrt{N_L/2}$, $P_L \equiv S_3/\sqrt{N_L/2}$, so that $[X_A, P_A]\approx i$ and $[X_L, P_L]\approx i$ (units of $\hbar$ ).   The Faraday interaction can then be expressed as, 
\begin{equation}
U_F = e^{-i \sqrt{\xi} P_A P_L},
\end{equation}  
where $\xi \equiv N_A N_L \chi^2/4= \rho \eta /9$  is the coupling strength. Here, $\rho=N_A (\sigma_0/A)$ is the characteristic resonant optical density and $\eta=N_L (\sigma_0/A) \left(\Gamma^2/4 \Delta^2\right)$ is the characteristic photon scattering probability per atom at detuning $\Delta$.  Note that $\rho$ and $\eta$ are defined with respect to a unit oscillator strength, with the Clebsch-Gordan coefficients appearing explicitly in the coupling strength.  In the HPA, the Faraday interaction displaces the $X$-quadratures of the spin and polarization subsystems, each by an amount proportional to the $P$-quadrature of the other, 
\begin{equation}
X_A^{out}=X_A^{in}+\sqrt{\xi} P_L^{in}, \; \; X_L^{out}=X_L^{in}+\sqrt{\xi} P_A^{in},
\end{equation}
and conserves the $P$-quadratures. Physically, the coupling strength $\xi$ quantifies the spin-polarization entanglement that results from stimulated emission of radiation by the atom ensemble into the probe mode.  This makes it a key parameter determining the performance of our atom-photon interface.

\begin{figure}
[t]\resizebox{7.5cm}{!}
{\includegraphics{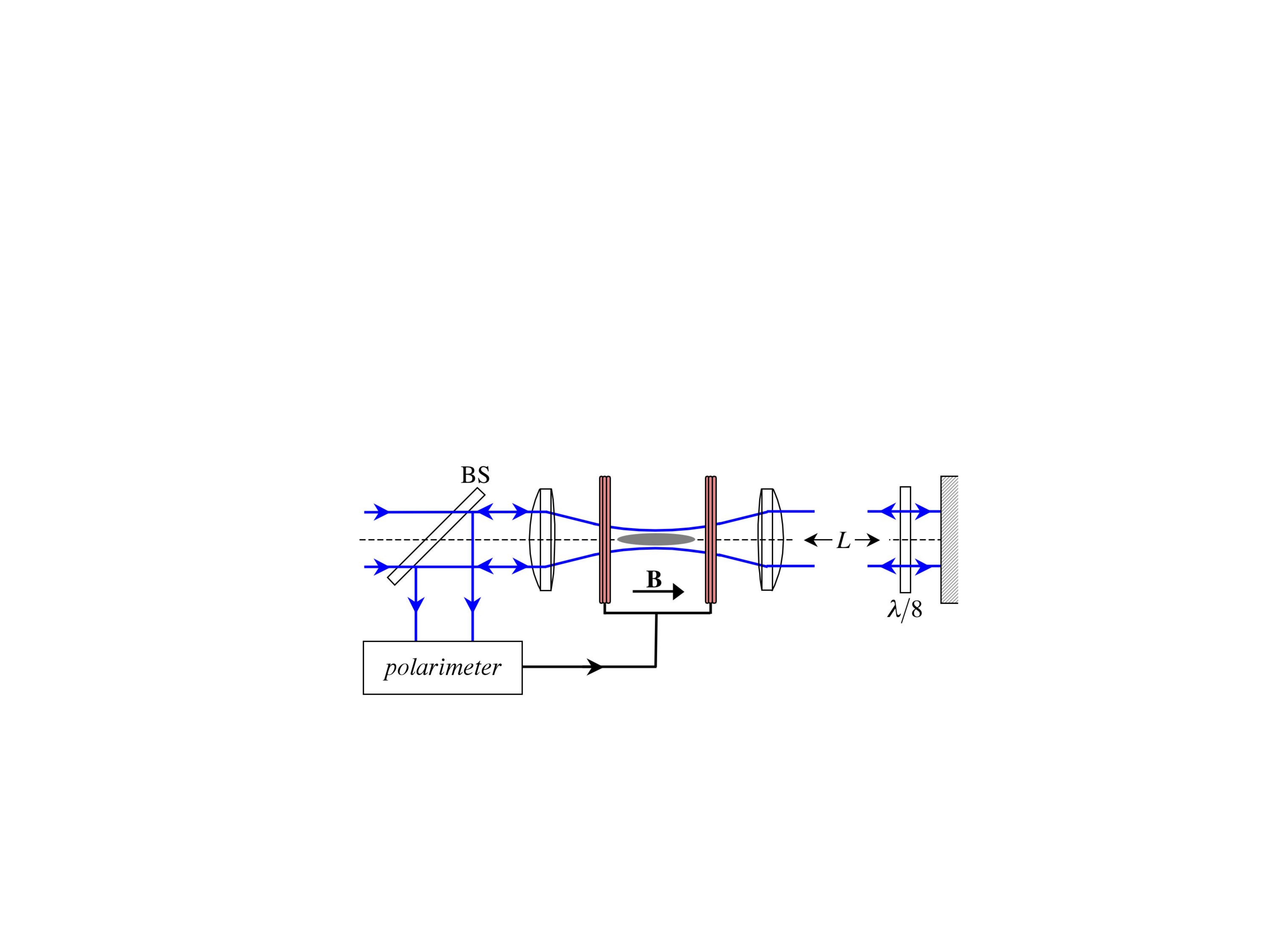}}
\caption{\label{fig:figure1}(Color online) Double-pass geometry for spin squeezing.  The probe beam undergoes Faraday rotation in the first pass and acts like a fictitious magnetic field during the second pass. A polarimeter and magnetic feedback controller removes spin-probe entanglement through quantum erasure, by measuring a complementary polarization observable and rotating the spin conditioned on the result. Short probe pulses and a long optical path length $L$ avoids standing wave effects.}
\end{figure}

Consider now the geometry shown in Fig.~\ref{fig:figure1}, consisting of a cigar-shaped ensemble of atoms coupled to a mode-matched, paraxial probe beam \cite{muller2005}. We take the initial polarization of the probe to be linear along $y$, and the initial state of the collective spin to be a coherent state along $x$. As the probe pulse passes through the ensemble along the $z$-axis, its polarization becomes correlated with quantum fluctuations in $J_z$, and a measurement of the Farady rotation corresponds to a QND measurement of $J_z$. Quantum backaction occurs and leads to spin squeezing when the signal from spin projection noise exceeds shot-noise in the polarimeter \cite{jessen2003}. In the limit of Gaussian statistics, one can show that the metrologically-defined squeezing parameter \cite{wineland1994} resulting from the measurement is $\zeta_{QND} \equiv N_A (\Delta J_z^2)_{QND} / \langle J_x \rangle^2 = (1+\xi)^{-1}$  \cite{hammer2008}. For large interactions, QND measurement leads to a scaling of the squeezing with coupling strength, $\zeta_{QND} \approx 1/\xi$.
	
Spin squeezing without measurement can be achieved via coherent feedback of the correlations created by the Faraday interaction, as outlined above \cite{takeuchi2005,vuletic2010}. The key is to employ a double-pass (DP) geometry, where the correlations created by Faraday interaction during the first pass are transformed into a fictitious magnetic field that rotates the spin by an amount proportional to its $J_z$ component during the second pass.  Quantitatively, the overall unitary transformation is a composition of a Faraday interaction, a polarization rotation by a waveplate, and a second Faraday interaction.  In the HPA this can be written as,
\begin{eqnarray}
U_{DP}=U_F e^{i \frac{\pi}{2} S_1} U_F \approx  e^{-i \sqrt {\xi} P_A \bar{P}_L} e^{-i \xi P_A^2/2} e^{i \frac{\pi}{2} a^\dagger a},  
\end{eqnarray}
where the Stokes bosonic operator for $x$-polarization in the HPA is $a=(X_L+i P_L)/\sqrt{2}$. Up to an initial overall rotation of the Stokes vector about the $S_1$ axis, the effect of the DP geometry is thus a nonlinear Òsingle-axis twistingÓ of the collective spin, $\propto J_z^2 \approx P_A^2$, which leads to spin squeezing \cite{kitagawa1993}.  In addition, $U_{DP}$ correlates the spin and polarization through a $\sqrt{2 \xi} P_A$ translation along the $45^\circ$  quadrature $\bar{X}_L=(X_L+P_L)/\sqrt{2}$ (generated by the conjugate observable $\bar{P}_L=(-X_L+P_L)/\sqrt{2}$).  For this map, one can show that the spin fluctuations along the optimal quadrature have a squeezing parameter given by $\zeta_{DP}(\theta_{min}) =1+\left(\xi^2+2 \xi \right)\left(1/2-\sqrt{1/2+(2+\xi)^{-2}}\right)   \Rightarrow \lim_{\xi \rightarrow \infty} 2/\xi.$  The DP protocol thus leads to the same $1/\xi$  scaling of the squeezing as the QND protocol, but with excess noise due to the residual entanglement between spin and light.

The excess noise seen in the DP geometry can in principle be eliminated by disentangling the spin and polarization degrees of freedom via a quantum eraser protocol \cite{scully1981}.  The key point is to erase ``which way" information carried by the probe by an appropriate projective measurement of its polarization, followed by a rotation on the spins that is conditioned on the measurement result.   To see this explicitly, consider the Heisenberg operator map generated by $U_{DP}$  that entangles the spin and $\pm 45^\circ$ polarization quadratures,
\begin{subequations}
\begin{align}
X_A^{out}=X_A^{in}+\xi P_A^{in}+\sqrt{2 \xi} \bar{X}_L^{in},\; \;  P_A^{out}=P_A^{in} \\
\bar{X}_L^{out}=-\bar{P}_L^{in}+\sqrt{2 \xi} P_A^{in}, \; \;  \bar{P}_L^{out}=\bar{X}_L^{in}
\end{align}
\end{subequations}
$P_A$  is a QND observable, whereas $X_A$ is displaced proportionally to $P_A$ as in a free particle evolution.   The additional coupling of $X_A^{out}$ to $\bar{X}_L^{in}$ represents the residual entanglement responsible for excess noise.  However, the conjugate observable $\bar{P}_L^{out}$ contains no information about the spin, and thus upon measuring this quadrature, we project the system to a random but known value of $\bar{X}_L^{in}$, distributed by a Gaussian according to the shot noise.  An additional displacement of $X_A$ (i.e., rotation of the spin around $z$) proportional to the measured value, $-\sqrt{2 \xi} \bar{P}_L^{out}$, removes the excess noise.  For a perfect quantum eraser, the spin is then mapped by a unitary transformation $U_{QE}=e^{-i \xi P_A^2 /2}$ , which geometrically ÒshearsÓ the initial coherent state uncertainty distribution along the $X_A$-axis, $X_A^{out}=X_A^{in}+\xi P_A^{in}$, $P_A^{out}=P_A^{in}$.  Under this transformation, squeezing occurs along the quadrature $\theta = \tan^{-1} (2/ \xi) /2 + \pi/2$, with a squeezing parameter, $\zeta_{QE}(\theta_{min})=1+\xi^2\left(1/2-\sqrt{1/2+\xi^{-2}}\right)  \Rightarrow \lim_{\xi \rightarrow \infty} 1/\xi^2$.  In contrast to the QND and DP protocols, the use of the quantum eraser thus allows a quadratic decrease in spin fluctuations with measurement strength.

\begin{figure*}
[t]\resizebox{15.0cm}{!}
{\includegraphics{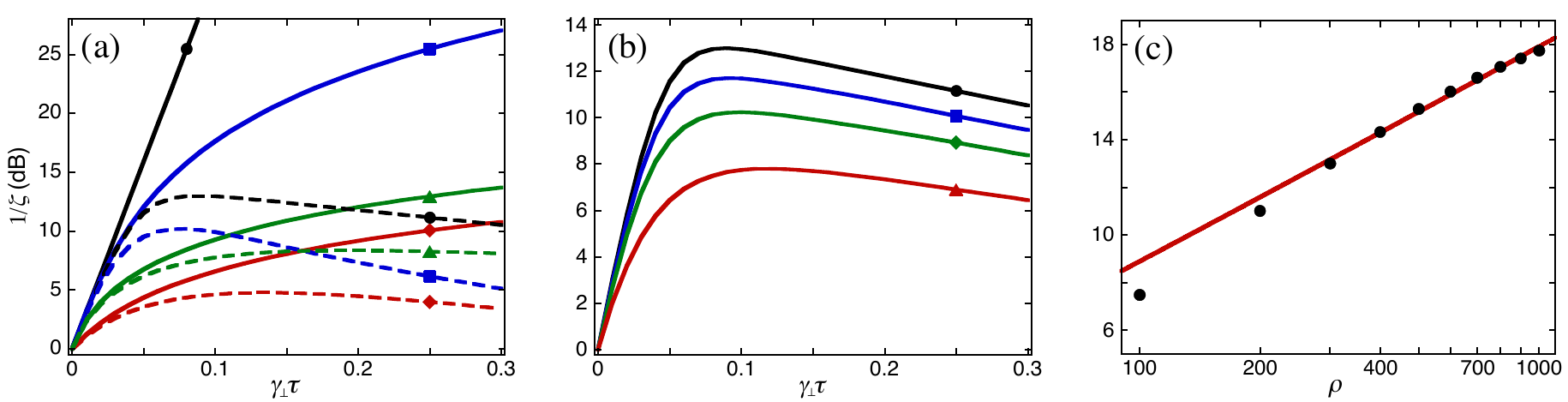}}
\caption{\label{fig:figure2} (Color online) (a) Performance of various spin squeezing protocols versus interaction time in a double pass, where $\gamma_{\perp}$ is the rate of decay of $J_z$, and $\gamma_{\perp} \tau = 4\eta/9$: QND (green/triangle), DP (red/diamond), QE (blue/square), PM (black/circle). Solid lines correspond to ideal protocols, dashed lines to models including photon scattering. (b) Performance of the PM protocol, in the presence of photon scattering and technical limitations imposed by optical loss and detector noise (in fractions of the probe shot noise): no loss and no noise (black/circle), 2\% loss and 1\% noise (blue/square), 6\% loss and 3\% noise (green/diamond), 20\% loss and 10\% noise (red/triangle).  In both (a) and (b) the optical depth is $\rho=300$.  (c) Peak squeezing versus $\rho$, including photon scattering but no additional loss or noise.  The line is a fit to the numerical data points for $\rho\ge300$.}
\end{figure*}

While the addition of a quantum eraser significantly improves the scaling of squeezing with $\xi$, further dramatic improvement results from the capacity for quantum control.  Squeezing arises from parametric instability and is associated with exponential shrinking of uncertainty with coupling strength \cite{walls2008}.  The unitary transformation in the QE protocol, $U_{QE}=e^{-i \xi P_A^2/2}$, corresponds to a combination of pure squeezing and rotation as is apparent by writing the spin quadratures in bosonic modes, $b=(X_A+i P_A)/\sqrt{2}$, so that
\begin{equation}
P_A^2 = - (b^2 +b^{\dagger 2})/2+b^\dagger b + 1/2.
\end{equation}
We thus see that the shearing evolution does not lead to parametric instability because it is not {\em phase matched}.  The first term is the generator of a Bogoliubov transformation, yielding pure squeezing with exponential growth, while the second term generates a residual rotation and thus a phase mismatch (the constant term is negligible).  We can achieve phase matching by canceling this spurious rotation, according to the Trotter formula, 
\begin{equation}
U_{PM}=\lim_{n \rightarrow \infty} (e^{i \frac{\xi}{2n}b^\dagger b} e^{-i \frac{\xi}{2n} P_A^2} )^n=e^{i \xi (b^2+b^{\dagger 2})/4}.
\label{trotter}
\end{equation}
This corresponds to alternate shearing interactions of strength $\xi /n$ and small rotations of the error ellipse about the spin polarization axis by an angle $\xi /2n$.  The phase-matched transformation, $U_{PM}$, is a pure squeezing unitary map with complex squeezing strength $\tilde{r}=-i \xi /2$.  Spin fluctuations are squeezed along the $-45^\circ$ quadrature at a rate that shrinks them exponentially, giving $\zeta_{PM}=e^{-\xi}$.  If achievable, such  exponential scaling will greatly enhance our ability to generate massive entanglement and perform nontrivial collective spin control.  The ideal (decoherence free) scaling of squeezing with $\xi$ for the various protocols is shown in Fig.~\ref{fig:figure2}

In a real-world implementation the various idealized protocols will suffer from imperfections.  The most fundamental of these is decoherence due to photon scattering into empty modes of the field, which will always accompany the desired collective scattering into the probe mode.  Exactly how such scattering damages the atomic spin squeezing is a subtle matter that depends on the specifics of atomic structure and atom-probe coupling \cite{saffman2009}.  A complete treatment of decoherence, including the full hyperfine structure of alkali atoms, is beyond the scope of this paper.  We instead consider a simplified (though not necessarily optimal) model of spin-1/2 atoms driven on a $S_{1/2} \rightarrow P_{1/2}$ transition.  Such a model is applicable to recent experiments with $^{171}$Yb \cite{takahashi2009}.  

To treat decoherence we make a Gaussian approximation for the statistics of all fluctuations, and employ the covariance matrix formalism as discussed by Madsen and M$\o$lmer \cite{madsen2004}.  The overall interaction is broken up in short pulses of duration $\tau/2n$, so as to apply alternate shearing pulses and phase matching rotations.  The Gaussian map follows as in \cite{madsen2004}, here modified to take advantage of the anisotropic nature of atom-photon scattering.  In particular, the spin component along the light polarization decays twice as fast as the perpendicular components. We choose the polarization to be along $y$ because this results in the least damage to the spin squeezing.  In a double pass with total interaction time $\tau$, $J_y$ decays at a rate defined by $\gamma_{\parallel}\tau =8\eta/9$ whereas $J_x$ and $J_z$ decay as $\gamma_{\perp}\tau=4\eta/9$.  Furthermore, for anisotropic noise the phase matching rotation given in Eq. (\ref{trotter}) is not optimal, and we therefore numerically optimize each individual rotation to produce the best squeezing after the next shearing pulse. The resulting spin squeezing as a function of the optical pumping probability is shown in Fig.~\ref{fig:figure2}, for a detuning $\Delta/\Gamma = 10^3$, large enough to render the atomic sample essentially transparent.  The unit-oscillator-strength optical density is $\rho=300$, an optimistic but not unreasonable extrapolation of the current values achieved in optical dipole traps \cite{mitchell2009}.  Under these conditions the model predicts a peak squeezing of 13 dB at $\gamma_{\perp}\tau \approx 0.08$.  

Technical imperfections will impose limitations on the protocol beyond those of photon scattering.  Optical loss between passes will occur due to reflections off optical elements in the beam path, which reduce the transfer of quantum correlations amongst the atoms. Moreover, perfect quantum erasure assumes a projective polarization measurement, which in practice is compromised by optical loss, finite quantum efficiency and technical noise in the polarimeter.  We treat the detector noise as additional Gaussian fluctuations in the measured value of $\bar{P}_L$, with variance $\sigma^2$ relative to the shot noise. Figure ~\ref{fig:figure2} shows how our protocol performs for different degrees of imperfection.  Strong squeezing of more than 10 dB is seen for an optical loss of $6\%$ and a detector noise level at $3\%$ of the probe shot noise, and a very respectable squeezing of 7 dB still occurs for $20\%$ loss and $10\%$ noise.

To understand how decoherence affects the exponential enhancement of squeezing it is useful to consider a simple model wherein optical pumping adds spin noise proportional to the number of photons scattered.  In that case the squeezing parameter is $\zeta \approx \zeta_{ideal} + c \eta$, where $\zeta_{ideal}$ is the squeezing in the absence of scattering, and  the constant $c$ quantifies the noise per scattered photon. For a given scattering rate this sets a maximum interaction time before optical pumping degrades the squeezing, and thus determines how the minimum value of $\zeta$ scales with $\rho$. For the three protocols we have considered, where squeezing variances in the absence of decoherence are, $\zeta_{QND}=1/\xi$, $\zeta_{QE}=1/\xi^2$,   $\zeta_{PM}=e^{-\xi}$, and $\xi \sim \rho\eta$, the peak squeezing scales as $\zeta_{QND}^{min}\sim \rho^{-1/2}$, $\zeta_{QE}^{min} \sim \rho^{-2/3}$, and $\zeta_{PM}^{min} \sim \left(a+b\log(\rho)\right)\rho^{-1}$, respectively.  The use of the quantum eraser and phase matching thus fundamentally changes how the achievable squeezing scales with optical density for a given noise model.   To further quantify the effectiveness of the various protocol, we numerically calculate the peak squeezing at the optimal value of $\eta$ as a function of $\rho$, as plotted in Fig.~\ref{fig:figure1}, and fit to the simple formula above in the limit of large $\rho$.  In the absence of of other technical noise, the fit of the phase-matched protocol gives a maximum squeezing that scales  as $\zeta_{PM}^{max}=\left(12.4+0.81\log\rho\right)/\rho$, yielding $\sim\!\!13$ dB of squeezing at a unit-oscillator $\rho$ of 300.

In summary, we have studied how one can employ the tools of quantum control to strongly enhance the spin squeezing of an atomic ensemble resulting from a QND light-shift interaction.  Through coherent feedback and a quantum-eraser protocol, we can implement a unitary nonlinear interaction on the collective spin and strongly amplify the squeezing through the technique of phase-matching. The achievable squeezing for a given optical density $\rho$ will depend on the competition between coherent atom-probe coupling and noise from photon scattering out of the probe mode. For a simple model of spin-1/2 atoms we have seen that phase matching leads to a fundamentally new scaling of the peak squeezing, $\zeta\sim1/\rho$ (plus logarithmic corrections), in contrast to the $\zeta\sim1/\sqrt{\rho}$ scaling that has so far been assumed \cite{hammer2008}.  The ultimate scaling in an experiment will depend strongly on the applicable noise model.  For example, in an idealized two-color scheme examined by Saffman {\em et al.} \cite{saffman2009} where the spontaneous scattering process respects the QND symmetry, no extra noise is added to the squeezed quadrature and squeezing degrades only due to decay of the mean spin vector.  In that case, photon scattering into other modes does not change the scaling of squeezing with $\rho$ as compared with the decoherence-free case.  For the single-pass QND protocol this would yield a $1/\rho$ scaling.  For our phase-matched protocol this would preserve the exponential scaling for much longer times, perhaps pushing the quantum fluctuations beyond the HPA, where curvature of the Bloch sphere leads to non-Gaussian states and more general quantum control \cite{chaudhury2007}.

\acknowledgments 
We thank Morgan Mitchell and Leigh Norris for helpful discussions.  This work was supported by NSF Grants PHY-0653599 and 0653631, and by ONR Grant N00014-05-1-420.

\bibliography{SqueezingBib}
 
\end{document}